\documentclass[reprint,pra,superscriptaddress]{revtex4-1}

\usepackage{amsmath}
\usepackage{amssymb}
\usepackage{enumitem}
\usepackage{graphicx}
\usepackage{dcolumn}
\usepackage{multirow}
\usepackage{color}
\newcommand{\Hamiltonian}{\mathcal{H}}

\DeclareMathOperator\Ci{Ci}

\newcommand{\units}[1]{\ensuremath{\mathrm{#1}}}

\newcommand{\bra}[1]{\ensuremath{\langle#1|}}
\newcommand{\ket}[1]{\ensuremath{|{#1}\rangle}}

\newcommand{\yc}[1]{\textcolor{black}{#1}}

\begin{document}

\title{High-fidelity single-qubit gates in a strongly driven quantum dot hybrid qubit with $1/f$ charge noise}

\author{Yuan-Chi Yang}
\email[]{yang339@wisc.edu}
\affiliation{Department of Physics, University of Wisconsin-Madison, Madison, Wisconsin, 53706, USA}

\author{S. N. Coppersmith}
\email[]{snc@physics.wisc.edu}
\affiliation{Department of Physics, University of Wisconsin-Madison, Madison, Wisconsin, 53706, USA}
\affiliation{School of Physics, University of New South Wales, Sydney, NSW 2052,
Australia}

\author{Mark Friesen}
\email[]{friesen@physics.wisc.edu}
\affiliation{Department of Physics, University of Wisconsin-Madison, Madison, Wisconsin, 53706, USA}

\date{\today}

\begin{abstract}
Semiconductor double quantum dot hybrid qubits are promising candidates for high-fidelity quantum computing. 
However, their performance is limited by charge noise, which is ubiquitous in solid-state devices, and phonon-induced dephasing.
Here we explore methods for improving the quantum operations of a hybrid qubit, using strong microwave driving to enable gate operations that are much faster than decoherence processes. 
Using numerical simulations and a theoretical method based on a cumulant expansion, we analyze qubit dynamics in the presence of $1/f$ charge noise, which forms the dominant decoherence mechanism in many solid-state devices.
We show that, while strong-driving effects and charge noise both reduce the quantum gate fidelity, 
simple pulse-shaping techniques effectively suppress the strong-driving effects.
Moreover, a broad AC sweet spot emerges when the detuning parameter and the tunneling coupling are driven simultaneously.
Taking into account phonon-mediated noise, we find that it should be possible to achieve $X_{\pi}$ gates with fidelities higher than $99.9\%$.  
\end{abstract}

\maketitle

\section{Introduction}
Semiconductor quantum dot qubits are promising platforms for quantum information processing.
These qubits are controlled by manipulating either electric voltages or magnetic fields, using DC pulses or microwave drives.
Microwave driving has several benefits for high-fidelity gates, including phase control of the rotation axis, amplitude control of the gate speed, protection against low-frequency noise, and the ability to perform the operations while keeping the qubit centered at a sweet spot.
Recently, high-fidelity resonantly driven single-qubit gates have been realized in several quantum dot architectures, including single-electron spin qubits~\cite{Zajaceaao5965,Watson2018,Yoneda2018}, singlet-triplet qubits~\cite{Nichol2017}, and quantum dot hybrid qubits~\cite{KimWardSimmonsEtAl2015}.

Quantum dot hybrid qubits can be operated fully electrically, allowing for fast microwave manipulations, while maintaining insensitivity to charge noise by operating at an extended sweet spot. 
However, gate fidelities are still limited by detuning charge noise~\cite{Thorgrimsson2017,PhysRevA.95.062321}.
An obvious strategy for outpacing noise effects is to increase gate speeds by applying a strong microwave drive, although this can also cause control errors. 
In a previous paper, we showed that, in the absence of noise, high fidelities can be achieved by accounting for the strong-driving-induced Bloch-Siegert shift of the qubit frequency and employing smoothed pulse envelopes to suppress fast oscillations and leakage~\cite{PhysRevA.95.062321}.
In this paper, we include the effects of charge noise with a $1/f$ spectrum. 
We find that, while AC driving reduces the direct effects of low-frequency noise ~\cite{10.1038/ncomms3337, PhysRevB.72.134519,Wong2016}, the interplay between strong driving and charge noise can also suppress the fidelity.
Indeed, when crosstalk effects dominate, increasing the driving amplitude is typically harmful for the gate performance.
Clarifying the effects of strong driving on decoherence is therefore critical for achieving high-fidelity gates.

In this work, we use numerical and analytical methods to investigate the crosstalk between strong driving and charge noise.
We show that it is possible to achieve high-fidelity single-qubit rotations in quantum dot hybrid qubits under these conditions.
We also find that the infidelity caused by decoherence can be greatly reduced by driving the detuning and tunnel coupling simultaneously and coherently.
In this way, we show that $X_{\pi}$ gates with fidelities higher than $99.9\%$ can be achieved over a broad range of control parameters.
We further include phonon decoherence phenomenologically and show that $X_{\pi}$ gate fidelities can still be larger than $99.9\%$, provided that the phonon decoherence time scale is larger than $2 \;\units{\mu s}$.

The paper is organized as follows. 
Section~\ref{Sec:QuantumDotHybridQubit} briefly introduces the double quantum dot hybrid qubit. 
In Sec.~\ref{Sec:Methods}, we provide an analytical description of the hybrid qubit dynamics, subject to  charge noise in the detuning parameter, and compare these results to numerical simulations. 
In Sec.~\ref{Sec: Fidelity}, we propose methods for improving the fidelity and show that $X_{\pi}$ gates with fidelities $>99.9\%$ are feasible, even in the presence of realistic levels of $1/f$ charge noise and phonon dephasing. 
We conclude in Sec.~\ref{Sec:Conclusion}.
Additional technical details of the calculations are provided in the Appendices.

\begin{figure*}[t]
\includegraphics[width=7in]{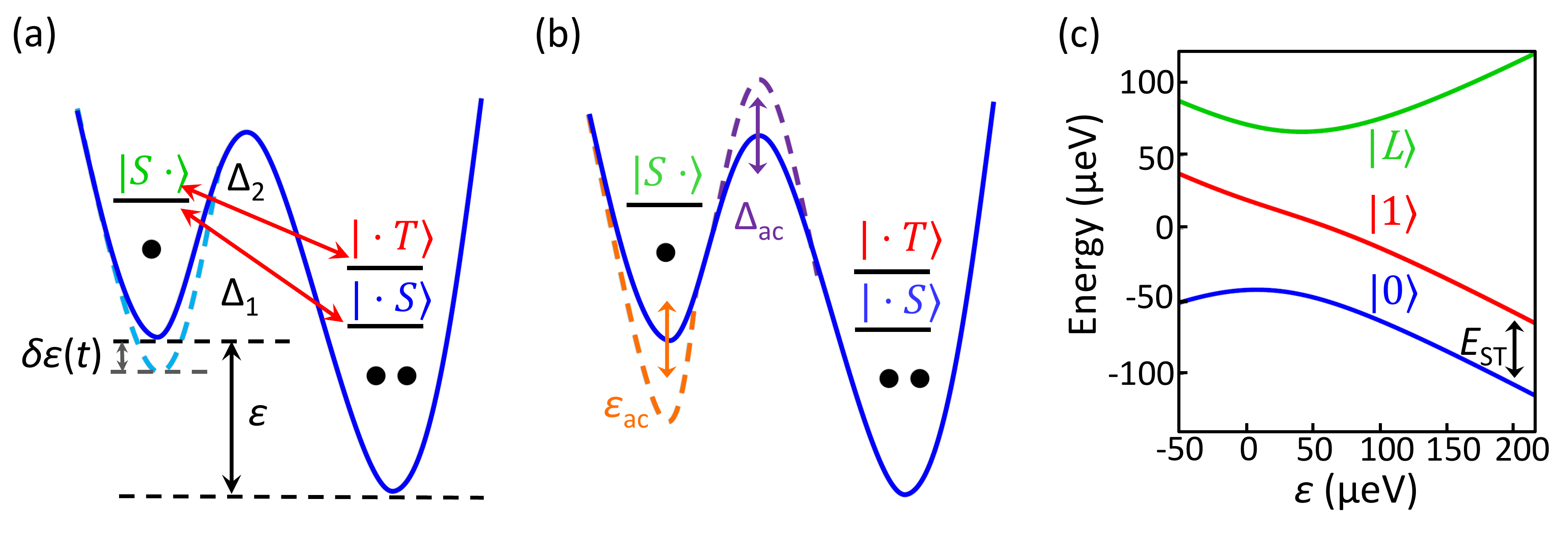}
\caption{
\label{Fig:Fig1}
Schematic confinement potential and energy levels of a quantum dot hybrid qubit.
(a), (b) A hybrid qubit is formed in a double quantum dot containing three electrons, as depicted here in the $(1,2)$ charge configuration.
In this arrangement, the low-energy basis states, shown in the right-hand dots, comprise singlet-like ($\ket{\cdot S}$) or triplet-like ($\ket{\cdot T}$) spin states, where $S$ and $T$ refer to the doubly occupied dot~\cite{Shi2012,Koh2012}.
The $(2,1)$ charge configuration has only one relevant low-energy basis state ($\ket{S\cdot}$), depicted in the left-hand dots.
(a) illustrates the Hubbard Hamiltonian parameters of the undriven system: the energy detuning between the left and right dots, $\varepsilon$, and the tunnel couplings, $\Delta_1$ and $\Delta_2$, between $\ket{S\cdot}$, and $\ket{\cdot S}$ or $\ket{\cdot T}$, respectively. 
The parameters $\Delta_1$, $\Delta_2$, and $\varepsilon$ are all controlled by voltages applied to the device top gates~\cite{KimWardSimmonsEtAl2015}.
Detuning fluctuations caused by charge noise, $\delta \varepsilon(t)$, are the dominant decoherence mechanism for this system~\cite{Thorgrimsson2017}.
(b) depicts the ac control of the detuning parameter $\varepsilon_\text{ac}$ and the tunnel couplings $\Delta_\text{ac}$, which are used to implement resonant gates. 
(c) A typical energy level diagram for a hybrid qubit, as a function of $\varepsilon$, showing the asymptotic energy splitting of the qubit states, $E_\text{ST}$.
Here, the lowest two levels (red and blue) correspond to the qubit subspace, while the highest level (green) corresponds to a leakage state.
(c) is obtained by diagonalizing Eq.~(\ref{Eq:HamiltonianQ}), assuming a realistic value of $E_\text{ST}/h = 12$~GHz~\cite{KimShiSimmonsEtAl2014}, and $\Delta_1=\Delta_2 = 0.7\,E_\text{ST}$.
}
\end{figure*}

\section{Theoretical Model \label{Sec:QuantumDotHybridQubit}}
We now present the full theoretical model used in this work for the quantum dot hybrid qubit.
The Hamiltonian is given by
\begin{equation}
\label{Eq:HybridQubitNoisy}
\Hamiltonian = \Hamiltonian_q+ \Hamiltonian_\text{ac} +\Hamiltonian_n  ,
\end{equation}
where the three components of the Hamiltonian are described below.

\textit{The Qubit Hamiltonian, $\Hamiltonian_q$.}
The quantum dot hybrid qubit is formed of three electrons in a double quantum dot, with total spin quantum numbers $S=1/2$ and $S_z=-1/2$~\cite{Shi2012,Koh2012}.
For the operating regime of interest, we consider the three-dimensional (3D) basis composed of $\ket{\cdot S}\equiv \ket{\downarrow\! S}$, 
$\ket{\cdot T}  \equiv \sqrt{1/3} \ket{\downarrow\! T_0} - \sqrt{2/3} \ket{\uparrow\! T_-}$, 
and $\ket{S\cdot} \equiv \ket{S\! \downarrow}$, 
where $\ket{\cdot}$ denotes a dot with one electron, the singlet state $|S\rangle=(\ket{\uparrow\downarrow}-\ket{\downarrow\uparrow})/\sqrt{2}$, and the triplet states $|T_0\rangle=(\ket{\uparrow\downarrow}+\ket{\downarrow\uparrow})/\sqrt{2}$ and $|T_-\rangle=\ket{\downarrow\downarrow}$ denote the spin states of dots with two electrons.
In this basis,  the qubit Hamiltonian is given by
\begin{equation}
\Hamiltonian_q = 
\begin{pmatrix}
-\frac{\varepsilon}{2} & 0 & \Delta_1 \\
0 & -\frac{\varepsilon}{2} + E_\text{ST} & -\Delta_2 \\
\Delta_1 & -\Delta_2 & \frac{\varepsilon}{2}
\end{pmatrix} ,
\label{Eq:HamiltonianQ}
\end{equation}
where $\varepsilon$ is the detuning parameter, corresponding to the energy difference between the dots, 
$E_\text{ST}$ approximately corresponds to the singlet-triplet energy splitting in the doubly occupied dot, 
and $\Delta_1$ ($\Delta_2$) are the tunnel couplings between the states $\ket{S\cdot}$ and $\ket{\cdot S}$ ($\ket{\cdot T}$). 
The various parameters are labelled in the schematic diagram shown in Fig.~\ref{Fig:Fig1}(a), and a typical energy diagram is shown in Fig.~\ref{Fig:Fig1}(c), where the two low-energy states $\ket{0}$ and $\ket{1}$ comprise the qubit, while the high-energy state $\ket{L}$ is a leakage state.
To simplify our analysis later, we transform Eq.~(\ref{Eq:HybridQubitNoisy}) to the energy basis, $\{|0\rangle, |1\rangle, |L\rangle\}$, yielding $\bar \Hamiltonian_q = \text{diag}[E_0,E_1,E_L] $, where the bar indicates the energy basis, and $\{ E_i\}$ are the corresponding energy eigenvalues.

\textit{The AC Drive,} $\Hamiltonian_\text{ac}$.
We consider two different schemes for AC driving \cite{Koh03122013}, as shown in Fig.~\ref{Fig:Fig1}(b). 
In the first case, we modulate the tunnel couplings as $\Delta_i = \Delta_{i0} + r_i \Delta_\text{ac}(t)$, where $i=1,2$ [purple arrow in Fig.~\ref{Fig:Fig1}(b)].
The AC drive $\Delta_\text{ac}$ is achieved by applying a microwave voltage signal to one of the device top-gates~\cite{KimWardSimmonsEtAl2015}.  
It is reasonable to assume that the same modulation drives both $\Delta_1$ and $\Delta_2$, although they may be affected differently, which we take into account through the variable $r_i$.
In the second case, we modulate the detuning as $\varepsilon = \varepsilon_{0} + \varepsilon_\text{ac}(t)$  [orange arrow in Fig.~\ref{Fig:Fig1}(b)].
We first consider rectangular pulses, $\Delta_\text{ac}(t) = A_{\Delta} \cos(\omega_d t)$ and $\varepsilon_\text{ac}(t) = A_{\varepsilon} \cos(\omega_d t)$, for which the pulse amplitudes $A_{\Delta}$  and $A_{\varepsilon}$ are piecewise constant in time, and $\omega_d$ is the driving angular frequency. 
The driving Hamiltonian expressed in the $\{|\cdot S\rangle, |\cdot T\rangle, |S \cdot\rangle\}$ basis can then be written as
\begin{equation}
\Hamiltonian_\text{ac} = V \! \cos(\omega_d t)=
\begin{pmatrix}
-\frac{A_{\varepsilon}}{2} & 0 & A_{\Delta} \\
0 & -\frac{A_{\varepsilon}}{2}  & -r A_{\Delta} \\
A_{\Delta} & -r A_{\Delta} & \frac{A_{\varepsilon}}{2}
\end{pmatrix} \!\cos(\omega_d t) ,
\label{Eq:Hamiltonianac}
\end{equation}
where we refer to $V$ as the driving matrix, and we simplify the expression by defining $r_1 = 1$ and $r_2 = r$. Note that we keep $r$ as a variable here for generality; however in our simulations, we choose $r=1$ for simplicity.

\begin{figure*}[t]
\includegraphics[width=7in]{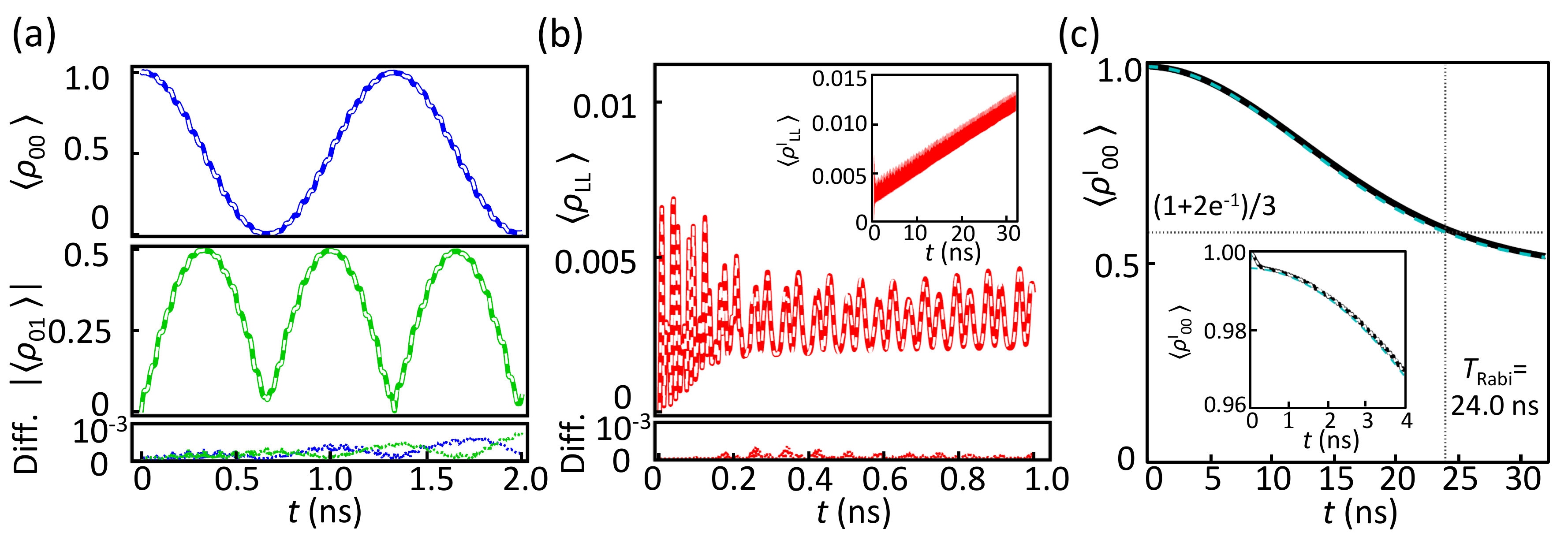}
\caption{
\label{Fig:Fig2}
Dynamics of a strongly driven quantum dot hybrid qubit in the presence of detuning fluctuations.
For all calculations, we assume the DC control parameters $\{\varepsilon, E_\text{ST}, \Delta_1,\Delta_2 \}/h = \{80, 12, 8.4, 8.4\}\; \units{GHz}$. 
Qubit evolutions are computed, taking $|0\rangle$ as the initial state, and applying a resonant AC signal of amplitude $A_{\Delta}/h = 3.5\; \units{GHz}$ to both tunnel couplings, with a ratio  of $r=1$ between the two driving amplitudes. 
The resulting density matrix is averaged over many realizations of $1/f$ charge noise, Eq.~(\ref{Eq:Spectrum One-Over-f}), assuming noise parameters $\sqrt{2\pi}c_{\varepsilon} = 2.38 \;\units{\mu eV}$, $\omega_{l}/2\pi = 1 \;\units{Hz}$, and $\omega_{h}/2\pi = 256 \;\units{GHz}$.
In all panels, the noise-averaged numerical simulations are plotted as solid lines, while corresponding analytical results, obtained up to second order in the cumulant expansion, Eq.~(\ref{Eq:AnalyticFormula2nd}), are plotted as white dashed lines.
In (a) and (b), the differences between numerical and analytical results are also plotted as dotted lines (bottom panels), indicating errors $<$0.1\%. 
(a) The average occupancy of the initial state, $\langle\rho_{00}\rangle$ (blue), and the off-diagonal element, $|\langle\rho_{01}\rangle|$ (green), computed in the lab frame. 
Here, the smooth sinusoidal envelope reflects Rabi oscillations, as consistent with the rotating wave approximation (RWA), while the fast modulations are caused by strong driving. 
(b) The average occupancy of the leakage state, $\langle\rho_{LL}\rangle$ (red), in the lab frame. 
The inset shows the accumulated leakage.  
In (a) and (b), the analytical results are seen to capture all the significant features of the simulations, including the fast oscillations and the asymptotic decay.  
(c) The asymptotic decay of the density matrix, $\langle\rho_{00}^I\rangle$, in the interaction frame. 
The inset shows a blown-up view at short times. 
Here, the analytical results correspond to the full solution of Eq.~(\ref{Eq:AnalyticFormula2nd}) (dashed white line, inset), and its asymptotic form, Eq.~(\ref{Eq:AsymptoticFormula}) (dashed cyan line).
Note that the fast oscillations, observed in the inset, are not captured in Eq.~(\ref{Eq:AsymptoticFormula}), but are accurately described in Eq.~(\ref{Eq:AnalyticFormula2nd}). 
As indicated, the Rabi decay time, $T_\text{Rabi} \simeq 24.0\; \units{ns}$, is determined according to the definition $\langle\rho_{00}^I(T_\text{Rabi})\rangle = (1+2e^{-1})/3$.
}
\end{figure*}

\textit{The Noise Hamiltonian, $\Hamiltonian_n$.}
The parameter $\varepsilon$ in Eq.~(\ref{Eq:HamiltonianQ}) represents the desired value of the detuning, and is controlled by voltages applied to the device top-gates.
However, charge noise within the device causes the detuning to fluctuate by $\delta \varepsilon(t)$, which represents the dominant source of decoherence for hybrid qubits~\cite{Thorgrimsson2017}.
The resulting noise Hamiltonian in the $\{|\cdot S\rangle, |\cdot T\rangle, |S \cdot\rangle\}$ basis is given by
\begin{equation}
\Hamiltonian_n = h_n \delta \varepsilon(t) = 
\begin{pmatrix}
-\frac{1}{2} & 0 & 0\\
0 & -\frac{1}{2}  & 0 \\
0 & 0 & \frac{1}{2}
\end{pmatrix} \delta \varepsilon(t) ,
\label{Eq:HamiltonianN}
\end{equation}
where $h_n$ is the dimensionless noise matrix.
The noise is characterized by the time correlation function $S(t_1-t_2) = \langle \delta\varepsilon(t_1) \delta\varepsilon(t_2) \rangle$, where the brackets denote an average over noise realizations and the noise is assumed to be stationary with zero mean ($\langle\delta\varepsilon\rangle=0$).
The corresponding noise power spectrum is given by $\tilde{S}(\omega) = \int_{-\infty}^{\infty} dt\,e^{i \omega t} S(t)$ \cite{RevModPhys.82.1155}.
In solid-state devices, the charge noise typically exhibits a $1/f$ power spectrum density. 
In this work, we adopt the following model for $1/f$ charge noise~\cite{nphys2688,RevModPhys.86.361}:
\begin{equation}
\label{Eq:Spectrum One-Over-f}
\tilde{S}(\omega) =  \left\{
  \begin{array}{cl}
   c_{\varepsilon}^2 \frac{2 \pi}{|\omega|} &  (\omega_l \leq |\omega | \leq \omega_h)\\
    0 &  (\text{otherwise}) 
  \end{array}
\right. ,
\end{equation}
where $c_{\varepsilon}$ is the noise amplitude, and $\omega_l$ ($\omega_h$) are the low (high) angular frequency cutoffs.
In the simulations described below, time sequences for $\delta \varepsilon(t)$ are obtained by first generating a white noise sequence, then scaling its Fourier transform by the spectral density function given in Eq.~(\ref{Eq:Spectrum One-Over-f}), following the procedure described in Refs.~\cite{Kawakami18102016,Yang:2019}, and Appendix~\ref{Sec:Simulation}.

\section{Theoretical Methods\label{Sec:Methods}}
In this section, we perform simulations and analytical calculations of a strongly driven hybrid qubit in the presence of charge noise, obtaining excellent agreement between the two methods. 
We also obtain simple analytical expressions for the decoherence rates of the density matrix in the long-time limit.

Figure~\ref{Fig:Fig2} shows the results of numerical simulations of the density matrix, using the methods described in Appendix~\ref{Sec:Simulation}, when the qubit is initialized into the $\ket{0}$ state.
Here and throughout this work, we choose fixed values of $E_\text{ST}$ and $\varepsilon$, as indicated in the figure caption, which are representative of quantum dot hybrid qubits~\cite{Thorgrimsson2017}.
The tunnel couplings are chosen such that $\Delta_1=\Delta_2=0.7E_\text{ST}$, which approximately corresponds to a second-order DC sweet spot for this system, in the limit of large $\varepsilon$~\cite{PhysRevA.95.062321}.
The AC driving parameter $A_\Delta$ is also chosen to be representative, experimentally; in later calculations, the AC parameters are allowed to vary.
The noise parameters are also specified in the figure caption, and the resulting density matrix is averaged over 10,000 realizations of $\delta \varepsilon(t)$.
For the results shown in Fig.~\ref{Fig:Fig2}(a), the qubit exhibits Rabi oscillations, as well as small-amplitude fast oscillations caused by leakage and the counter-rotating terms in the qubit evolution, as previously explained in Ref.~\cite{PhysRevA.95.062321}.
The leakage component of the density matrix also exhibits fast oscillations, as shown in Fig.~\ref{Fig:Fig2}(b).
However, distinct from Ref.~\cite{PhysRevA.95.062321}, the Rabi oscillations here decay due to charge noise.
Moreover, the leakage component accumulates on longer time scales, as seen in the inset of Fig.~\ref{Fig:Fig2}(b).

To differentiate between the effects of decoherence and strong driving, we recompute the noise-averaged density matrix after first transforming to the interaction frame, defined as $\rho^I = U_0^{\dagger} \rho U_0$, where $\rho$ is expressed in the energy basis, and $U_0$ is the time-evolution operator for the energy frame, including strong-driving dynamics but not charge noise, as derived in Appendix~\ref{Sec:AnalyticsNoNoise}.
In this interaction frame, noise-free dynamics simply correspond to $\rho^I\!(t)=\text{const.}$ 
However, the numerical results in Fig.~\ref{Fig:Fig2}(c) decay over time, due to the presence of charge noise.
Here, short-time behavior is plotted in the inset, where a careful examination shows that small-amplitude, high-frequency oscillations still persist, as an example of the crosstalk between noise and strong driving.
This particular effect can be classified as `non-Markovian,' as discussed below.

We now obtain analytical expressions for the dynamics of strongly driven hybrid qubits in the presence of charge noise using a technique developed for quantum dot charge qubits~\cite{Yang:2019}, which is based on a cumulant expansion. 
The method is summarized as follows, with further details provided in Appendix~\ref{Sec:AnalyticalCalculation}.
First, we calculate the qubit evolution in the interaction frame, as governed by the equation $i \hbar \, d\rho^I\!/dt = \delta \varepsilon(t) \mathcal{L} \rho^I$, where $\mathcal{L} \rho^I \equiv [h_n^I,\rho^I]$, $h_n^I = U_0^{\dagger} \bar h_n U_0$, and $\bar h_n$ is expressed in the energy basis. 
To determine $h_n^I$, we follow Ref.~\cite{PhysRevA.95.062321} in expanding $U_0$ in powers of the small parameter $\gamma \sim V/\hbar \omega_d$, obtaining $U_0 = U_0^{(0)} + U_0^{(1)} + U_0^{(2)} + \cdots$, where $U_0^{(n)}\propto\gamma^n$.
We then perform a noise average of $\rho^I$, similar to Ref.~\cite{Yang:2019}, adopting  $\delta \varepsilon/V$ as a small parameter in the cumulant expansion~\cite{doi:10.1143/JPSJ.17.1100}, obtaining
\begin{equation}
\label{Eq:AnalyticFormula2nd}
\langle \rho^I(t) \rangle =e^{ -\frac{1}{\hbar^2}  \int\limits_0^t dt_1 \int\limits_0^{t_1} dt_2 \mathcal{L}(t_1)  \mathcal{L}(t_2) S(t_1-t_2) } \rho^I(0) 
\end{equation}
at $\mathcal{O}[(\delta \varepsilon/V)^2]$.
To simplify the calculation, we expand our result in terms of Gell-Mann matrices $\{\lambda_i\}_{i=1,\cdots,8}$, as defined in Appendix~\ref{Sec:GellmanMatrix}, obtaining 
$\langle\rho^I\rangle = I_3/3 +\sum_{i=1}^8 \vec{r}^I_i \lambda_i/2$, 
where $I_3$ is the $3 \times 3$ identity matrix and $\vec{r}^I$ is a generalized, 8D Bloch vector.
The noise matrix in the interaction frame can also be expressed as $h_n^I = \sum_{i=1}^8 h_{n,i}^I \lambda_i$, and the Bloch vector can be rewritten as
\begin{equation}
\vec{r}^I(t) = \exp [K(t) ] \vec{r}^I(0), \label{Eq:rI(t)}
\end{equation}
where $K(t)$ is an $8 \times 8$ matrix. 
Expanding $h_{n,i}^I(t)$ in a Fourier series, $h_{n,i}^I(t) =\sum_{\omega}\alpha_{i,\omega} e^{i\omega t}$, we obtain
\begin{multline}
\label{Eq:K(t)}
[K(t)]_{ij} \\
= -\frac{1}{\hbar^2} \sum\limits_{\omega_1, \omega_2} \sum\limits_{k,l,m = 1}^{8}
\alpha_{k,\omega_1}\alpha_{l,\omega_2}  T_{k m}^{(i)}T_{lj}^{(m)}  I(t,\omega_1,\omega_2) ,
\end{multline} 
where $T_{ij}^{(k)}$ is a structure constant defined as $[\lambda_i, \lambda_j] \equiv \sum_{k=1}^8 T_{ij}^{(k)} \lambda_k$, and
\begin{equation}
I(t,\omega_1,\omega_2)  \equiv \int_0^t dt_1 \int_0^{t_1} dt_2 e^{i\omega_1 t_1} e^{i\omega_2 t_2} S(t_1-t_2) .
\label{Eq:I(t)}
\end{equation}

The physics of the noisy qubit dynamics is encoded in $K(t)$.
We can characterize different types of behavior by decomposing $K(t)$ into a sum of pure-dephasing ($K_{\varphi}$), Markovian ($K_\text{M}$),  and non-Markovian-non-dephasing terms ($K_{\text{nMn}\varphi}$)~\cite{Yang:2019}. 
Note that pure dephasing is defined, here, with respect to the interaction frame, 
and is associated with the integral $I(t,\omega_1\text{=}0,\omega_2$=$0) \sim t^2\ln(1/\omega_l t)$~\cite{PhysRevLett.88.228304,PhysRevLett.93.267007, PhysRevB.72.134519}.
The Markovian terms induce characteristic exponential decay ($e^{-\Gamma t}$), which is associated with the Markovian approximation and the integral $\text{Re}[I(t,\omega,-\omega)]$.
The non-Markovian-non-dephasing terms describe any additional effects, including \yc{the small-amplitude, high-frequency oscillations observed in Fig.~\ref{Fig:Fig2}(c)}. 

The formalism described above allows us to compare and contrast various decoherence mechanisms within a common framework.
We now describe the main results.
In Fig.~\ref{Fig:Fig2}, our analytical results are plotted as white dashed lines, for comparison with the simulations.
Figures~\ref{Fig:Fig2}(a) and \ref{Fig:Fig2}(b) demonstrate that the analytical calculations can accurately describe the main features of the simulations, including decoherence, with deviations between $\langle \rho_{00} \rangle$, $\langle \rho_{01} \rangle$, and $\langle \rho_{LL} \rangle$ and their simulated values falling below $10^{-3}$ in all cases (bottom panels).  

We can also obtain simple, approximate expressions for the long-time dynamics, as described in Appendix~\ref{Sec:AnalyticsNoise}.
For the initial state $\rho(t=0) = |0\rangle\!\langle0|$, we obtain the leading order behavior~\footnote{Note that we use slightly different notation here, as compared to Ref.~\cite{Yang:2019}. In the latter, $\Gamma_\varphi$ had units of $\text{(sec)}^{-2}$. Here, $\Gamma_\varphi$ has units of $\text{(sec)}^{-1}$, and is a true rate.}
\begin{equation}
\label{Eq:AsymptoticFormula}
\langle\rho^I_{00}\rangle(t) 
= \frac{1}{3} + \left(\frac{1}{2}-\frac{a}{6}\right)e^{-\Gamma_y t - 2 \Gamma_{\varphi}^2\varphi(t)} +\frac{1}{6} (1-a)e^{-\Gamma_L t},
\end{equation}
where 
\begin{widetext}
\begin{gather*}
\Gamma_y =  \left(\frac{1}{\hbar}\right)^2\left\{\frac{\bar h_{n,1}^2}{4}\left[\tilde S(\omega + \Omega)+\tilde S(\omega - \Omega) + 4 \tilde S(\omega )\right] 
 + \frac{\bar h_{n,5}^2}{4} \left[ \tilde S \left( \frac{E_L-E_0}{\hbar}
  + \frac{\Omega}{2}\right)+\tilde S \left( \frac{E_L-E_0}{\hbar}- \frac{\Omega}{2}\right) \right] 
  \right. \\ \left. \hspace{2in}
+ \frac{\bar h_{n,7}^2}{4}\left[\tilde S\left(\frac{E_L-E_1}{\hbar} + \frac{\Omega}{2}\right)+\tilde S\left(\frac{E_L-E_1}{\hbar}- \frac{\Omega}{2}\right)\right]\right\}, \\ 
\Gamma_L =  \left(\frac{1}{\hbar}\right)^2\left\{\frac{3 \bar h_{n,5}^2}{4}\left[\tilde S\left(\frac{E_L-E_0}{\hbar} + \frac{\Omega}{2}\right)+\tilde S\left(\frac{E_L-E_0}{\hbar}- \frac{\Omega}{2}\right)\right] 
\right. \\ \left. \hspace{2in}
+   \frac{3 \bar h_{n,7}^2}{4}\left[\tilde S\left(\frac{E_L-E_1}{\hbar} + \frac{\Omega}{2}\right)+\tilde S\left(\frac{E_L-E_1}{\hbar}- \frac{\Omega}{2}\right)\right]\right\} , \\ 
\Gamma_{\varphi} ^2= \left(\frac{c_{\varepsilon}}{\hbar}\right)^2 \left(\frac{d f_\text{Rabi}}{d\varepsilon}\right)^2, \\ 
\varphi(t) = t^2 \left[\log(1/\omega_l t)) -\gamma_E+ 3/2 \right], \\  \vspace{1in}
a = \frac{3 (E_L-E_0)^2 \bar V_{0L}^2}{(E_L-E_0-hf)^2(E_L-E_0 + hf)^2}  
+ \frac{3 (E_L-E_1)^2 \bar V_{1L}^2}{(E_L-E_1 -hf)^2(E_L+E_1 + hf)^2} .
\end{gather*}
\end{widetext}
Here, $\Gamma_y$ and $\Gamma_L$ represent decoherence rates associated with the Markovian approximation ($\Gamma_L$ specifically describes leakage processes),
$\Gamma_\varphi$ is the pure-dephasing rate, $a$ describes the asymptotic occupation of the leakage state, and we define $\bar h_n = \sum_{i=1}^8 \bar h_{n,i} \lambda_i$.
Note that the terms $\varphi$ and $a$ are induced by strong driving and are therefore responsible for the crosstalk with decoherence processes.
Equation~(\ref{Eq:AsymptoticFormula}) [plotted as a cyan dashed line in Fig.~\ref{Fig:Fig2}(c)] accurately describes the coarse features of the simulation, except at extremely short times (inset), where it incorrectly assigns a non-zero leakage.
A more accurate description of the short-time behavior, and other fine-scale features, requires retaining the full analytical expressions, as shown in the inset of Fig.~\ref{Fig:Fig2}(c) (dashed white line).
Since $\varphi(t)\sim t^2\log(t)$ dominates the decay of the density matrix for typical gate times (see below), it suggests that reducing $\Gamma_{\varphi}$ would significantly improve qubit performance.

\begin{figure*}[t]
\includegraphics[width=7in]{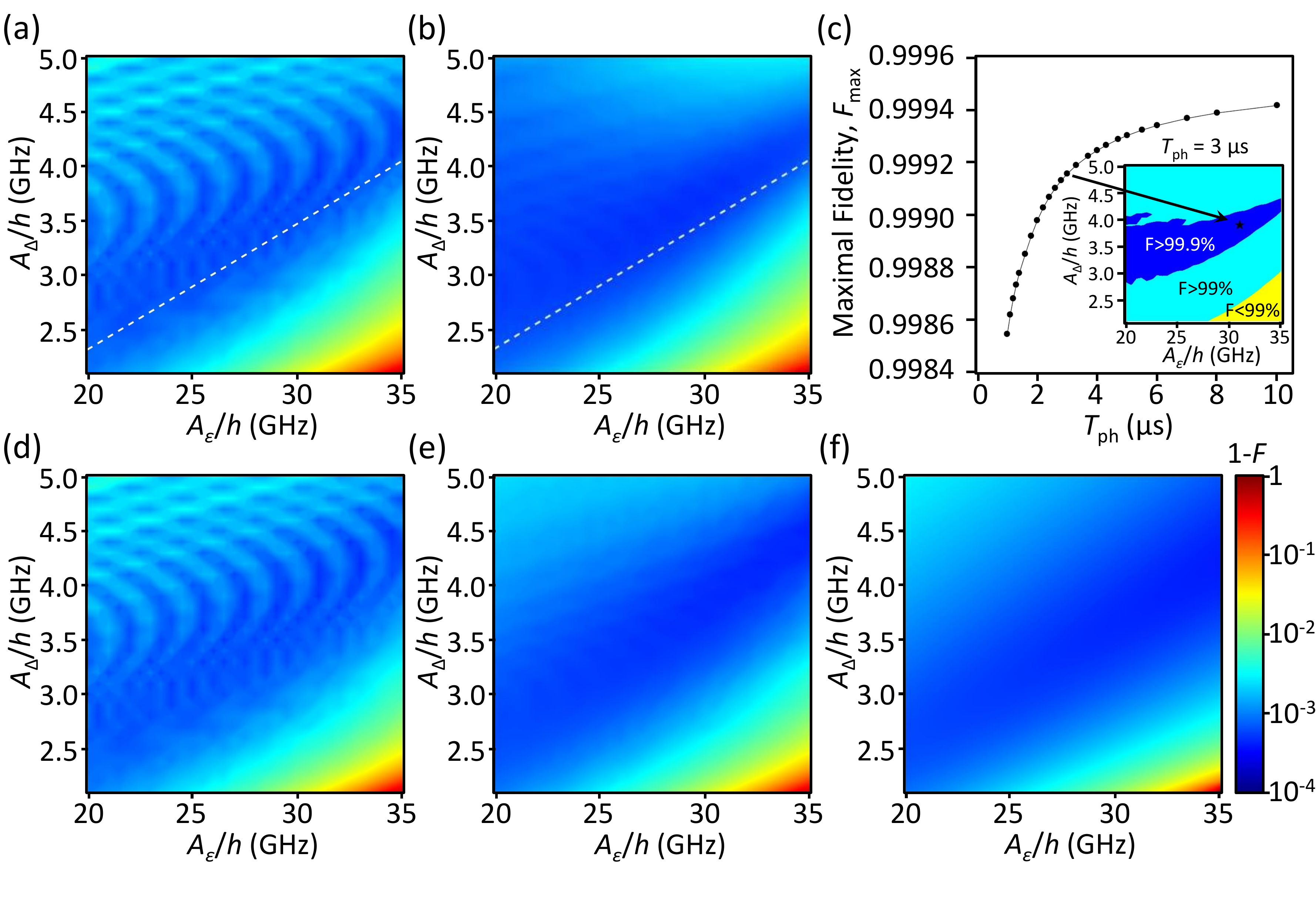}
\caption{
\label{Fig:Fig3}
$X_{\pi}$ gate fidelities of a strongly driven quantum dot hybrid qubit in the presence of $1/f$ detuning noise. 
Plots show the dependence of the fidelity on the detuning driving amplitude, $A_{\epsilon}$, and tunnel coupling driving amplitude, $A_{\Delta}$, using simulation parameters that are the same as Fig.~\ref{Fig:Fig2}.
Panels (a)-(d) are computed in the laboratory frame, while panels (e) and (f) are computed in the interaction frame, which represents an upper bound on the fidelity in the laboratory frame, since in this case, strong-driving effects are viewed as part of the coherent evolution in the interaction frame.
(a), (b) Simulation results obtained for (a) rectangular pulse envelopes, or (b) smoothed-rectangular pulse envelopes, with rise and fall times of $0.83\,\units{ns}$. 
For the rectangular pulse envelope, the fidelity exhibits fringes due to strong driving.
The dashed white line corresponds to an AC sweet spot, where $d f_\text{Rabi}/d\varepsilon=0$; in a broad region near this line, we observe fidelities $>$99.9\%.
For the smoothed-rectangular pulse envelope, the fidelity fringes are suppressed, and the quality of the AC sweet spot is improved.
In (c), the inset shows simulation results, similar to (a) and (b), where we phenomenologically include the effects of phonon-induced dephasing according to Eq.~(\ref{Eq:FidelityPhonon}), with $T_\text{ph}=3$~$\mu$s.
Here, we observed broad regions with fidelities $>$99.9\%, and
the location of the fidelity maximum is indicated with a star.
In the main panel, we plot a series of fidelity maxima, obtained in the same manner, as a function of $T_\text{ph}$, where the starred point corresponds to the inset.
(d) Analytical results obtained from Eq.~(\ref{Eq:AnalyticFormula2nd}), keeping expansion terms up to ${\cal O}[(V/\hbar \omega_d)^2]$.
The simulation results in (a) and analytical results in (d) are nearly identical, except in the lower-right portion of the plots, where higher-order terms in the expansion are nonnegligible.
(e), (f) Here, in the interaction frame, the main fringes due to strong driving are absent, and any suppression of the fidelity can be attributed to charge noise, or crosstalk between charge noise and strong-driving effects.
In (e) we plot the full analytical results based on Eq.~(\ref{Eq:AnalyticFormula2nd}), expanding up to ${\cal O}[(V/\hbar \omega_d)^2]$.
In (f), we plot the asymptotic results based on Eq.~(\ref{Eq:FidelityIAsymptotic}). 
We see that the simpler asymptotic results accurately reproduce the main features of the full analytical results, allowing us to determine the location of the AC sweet spot straightforwardly. }
\end{figure*}

\section{High-fidelity $X_{\pi}$ Gates \label{Sec: Fidelity}}
In this section, we first obtain an analytical expression for the fidelity at the asymptotic time scales relevant for quantum gates.
We then demonstrate that, by driving the detuning and tunnel coupling simultaneously, in a specific ratio, pure dephasing can be fully suppressed, $\Gamma_{\varphi} =0 $, in principle enabling significant improvements in the fidelity.
Finally we show, via simulations and analytical calculations, that in the presence of charge noise and strong driving effects, the fidelity of an $X_{\pi}$ gate can be higher than $99.9\%$.

\subsection{Asymptotic Fidelity Results}
We define the asympotic time regime as $1/\omega_l \gg t/(2\pi)\sim 1/\Omega \gg 1/\omega \gg 1/\omega_h$, where $\omega$ represents any angular frequency associated with resonant driving, other than the Rabi frequency $\Omega$, which is the slowest frequency in the system.
In this regime, Eq.~(\ref{Eq:AsymptoticFormula}) approximates the noise-averaged, driven density matrix in the interaction frame.
As outlined in Appendices~\ref{Sec:AnalyticsNoise} and \ref{Sec:ProcessFidelity}, we can use this to compute the process fidelity as a function of time, obtaining the simple expression
\begin{equation}
\label{Eq:FidelityIAsymptotic}
F^I = 1-\frac{1}{12}[3 \Gamma_x + 6 \Gamma_y + \Gamma_L ]t -\frac{1}{2}\Gamma_{\varphi}^2\varphi(t)-\frac{1}{3}a ,
\end{equation}
where 
\begin{multline*}
\Gamma_x =  \left(\frac{1}{\hbar}\right)^2\left\{\frac{\bar h_{n,1}^2}{2}\left[\tilde S(\omega + \Omega)+\tilde S(\omega - \Omega)\right] \right. \\
+ \frac{\bar h_{n,5}^2}{4}\left[\tilde S\left(\frac{E_L-E_0}{\hbar} + \frac{\Omega}{2}\right)+\tilde S\left(\frac{E_L-E_0}{\hbar}- \frac{\Omega}{2}\right)\right]  \\ 
\left. + \frac{\bar h_{n,7}^2}{4}\left[\tilde S\left(\frac{E_L-E_1}{\hbar} + \frac{\Omega}{2}\right)+\tilde S\left(\frac{E_L-E_1}{\hbar}- \frac{\Omega}{2}\right)\right]\right\}
\end{multline*} 
is also associated with the Markovian approximation.
For typical system parameters, we find that the $\Gamma_\varphi$ term in Eq.~(\ref{Eq:FidelityIAsymptotic}) dominates the infidelity.
Suppressing this contribution requires suppressing $df_\text{Rabi}/d\varepsilon$, or equivalently, $d\bar V_{01}/d\varepsilon$, where $\bar V_{ij}$ refers to a particular component of the $V$ matrix in the energy basis.
Remarkably, we find that it is possible to satisfy this condition exactly,  $df_\text{Rabi}/d\varepsilon=d\bar V_{01}/d\varepsilon=0$, over a continuous range of driving parameters $A_\varepsilon$ and $A_\Delta$, as consistent with an AC sweet spot~\cite{2018arXiv180701310D}.
In the present case, it is easy to show that \yc{$\bar V_{0L}$, $\bar V_{1L}$, and thus $a$ are suppressed at such tunings}, further improving the fidelity.

We illustrate the effectiveness of the AC sweet spot technique by simulating the $X_{\pi}$ gate fidelity for a 2D sweep over the driving amplitudes $A_{\varepsilon}$ and $A_{\Delta}$, as shown in Fig~\ref{Fig:Fig3}(a). 
Here, we compute the process fidelity, as described in Appendix~E, by comparing the noise-averaged simulation results to an ideal $X_\pi$ rotation.
The location of the AC sweet spot is determined numerically and plotted as a white dashed line. 
Near this line, we observe a broad region with gate fidelities greater than $99.9 \%$. 
Further away from the line, in the upper-left portion of the plot, we observe fast oscillations, which arise due to strong driving (e.g., counter-rotating terms; see the discussion below).
In the lower-right portion of the plot, we observe a region of low fidelity; here, the two driving fields interfere to produce a slow gate that is strongly affected by charge noise.
To demonstrate the effectiveness of our theoretical formalism, we also compute the $X_{\pi}$ gate fidelity theoretically, using Eq.~(\ref{Eq:AnalyticFormula2nd}).
These results, shown in Fig.~\ref{Fig:Fig3}(d), quantitatively reproduce all the features of the simulations. 

\subsection{Suppressing Strong-Driving Effects}

In the previous section, we noted that the oscillations observed in the top-left corner of Figs.~\ref{Fig:Fig3}(a) and \ref{Fig:Fig3}(d) are caused by strong driving.
To substantiate this claim, we determine the $X_\pi$ gate fidelity in the interaction frame, where the `ideal' density matrix, used to compute $F^I$, encompasses the full coherent evolution, including strong driving effects.
The results, plotted in Fig.~\ref{Fig:Fig3}(e), exhibit strongly suppressed oscillations, indicating that the oscillations in Fig.~\ref{Fig:Fig3}(d) indeed arise from strong driving.
Closer inspection of Fig.~\ref{Fig:Fig3}(e) reveals weak, residual oscillations, which arise at higher order in the strong-driving $\gamma$ expansion, and are a manifestation of the crosstalk between strong-driving effects and charge noise.
In Fig.~\ref{Fig:Fig3}(f), we also plot the simplified form of $F^I$ obtained in Eq.~(\ref{Eq:FidelityIAsymptotic}), which contains strong-driving corrections, including as the Bloch-Siegert shift of the resonant frequency, but no strong-driving fast oscillations.
The results reproduce the essential features in Fig.~\ref{Fig:Fig3}(e), explaining why this result can accurately predict the position of the AC sweet-spot line, as plotted in Fig.~\ref{Fig:Fig3}(a). 

In experiments, it is easy to account for the Bloch-Siegert shift by recalibrating the resonance frequency.
However, avoiding the fidelity oscillations in Fig.~\ref{Fig:Fig3}(a) may require careful tuning.
A simpler approach is to incorporate a smooth pulse envelope $p(t)$ into the driving term $\bar \Hamiltonian_\text{ac} = \bar V p(t) \cos(\omega_d t)$~\cite{PhysRevLett.115.133601,PhysRevA.94.032323}, as opposed to the rectangular pulse envelope used in Fig.~\ref{Fig:Fig3}(a).
To investigate this possibility, we consider a ``smoothed-rectangular" pulse envelope, defined as~\cite{PhysRevA.95.062321}
\begin{equation}
p(t) = \left\{ 
\begin{array}{cc} \vspace{0.03in}
\frac{t_g[1- \cos(\pi t/t_r) ]}{2(t_g-t_r)} &  (0\leq t \leq t_r), \\ \vspace{0.03in}
\frac{t_g}{t_g-t_r} & ( t_r< t < t_g - t_r), \\
\frac{t_g[1+ \cos(\pi [t-t_g+t_r]/t_r) ]}{2(t_g-t_r)} \hspace{0.2in} &  (t_g-t_r \leq t \leq t_g), 
\end{array}
\right.
\label{Eq:SmoothRectangular}
\end{equation} 
where $t_g$ is the pulse width, and we choose a rise time of $t_r = h/E_{ST} \sim 0.83\,\units{ns}$, to give a pulse that is sufficiently adiabatic.
Figure~\ref{Fig:Fig3}(b) shows the improved simulation results for the gate fidelity in the presence of charge noise, using this pulse form.
In comparison with Fig.~\ref{Fig:Fig3}(a), we now observe a suppression of strong-driving effects and a broad regime with gate fidelities $>$$99.9\%$. 
In fact, the gate fidelities obtained with the smoothed rectangular pulse are nearly identical to those in Fig.~\ref{Fig:Fig3}(e), which do not include direct strong-driving effects, therefore representing an upper bound.
This suggests that charge noise, rather than strong driving, limits the fidelity in this case.

\subsection{Phenomenological Treatment of Phonons}
A full treatment of phonons is outside the scope of this work; however, the following considerations allow us to estimate their effect.  
Dephasing is known to be the greatest threat from phonons for hybrid qubits, occurring on time scales of order microseconds~\cite{PhysRevB.86.035302}.
We account for this process here by phenomenologically expressing the total fidelity as
\begin{equation}
\label{Eq:FidelityPhonon}
F_\text{ch+ph} \approx F - t_g/T_\text{ph},
\end{equation}
as appropriate for Markovian processes at short times.
Here, $F$ is the fidelity obtained in previous sections, describing the effects of charge noise, $t_g$ is the gate time, and $T_\text{ph}$ is the phonon decoherence time.
As apparent in this expression, phonon dephasing effects are less effective for shorter gates.

In the inset of Fig.~\ref{Fig:Fig3}(c), we plot the results of a typical $X_\pi$ gate fidelity calculation  based on Eq.~(\ref{Eq:FidelityPhonon}), including both charge noise and phonon effects, and assuming smoothed rectangular pulse envelopes.
While $T_\text{ph} = 3 \; \units{\mu s}$ is held constant throughout this plot, we note that $t_g$ is a function of both $A_\varepsilon$ and $A_\Delta$.
Here we observe a relatively large region with fidelities $>$$99.9\%$, with the maximum fidelity occurring at the spot marked with a star.
Repeating this calculation for a range of $T_\text{ph}$, we obtain the fidelity maxima shown in the main panel of Fig.~\ref{Fig:Fig3}(c).
Generally, we conclude that fidelities $>$$99.9\%$ can be achieved when \yc{$T_\text{ph}> 2 \,\units{\mu s}$}.

\section{Summary and Conclusions \label{Sec:Conclusion}}
We have studied the dynamics of a strongly driven double quantum dot hybrid qubit in the presence of $1/f$ detuning charge noise, both analytically and numerically.
Our analytical results accurately reproduce the numerical simulations, and therefore provide insight into the dependence of the fidelity on the experimental parameters. 
In particular, the asymptotic fidelity in Eq.~(\ref{Eq:FidelityIAsymptotic}) is quite accurate for typical gate times, and can therefore be used to design high-fidelity gate protocols.
Using these results, we have shown that high-fidelity $X_{\pi}$ gates can be achieved by simultaneously and coherently driving the detuning and tunnel coupling, and that unwanted fast oscillations caused by strong driving can be suppressed by using smoothed rectangular pulse envelopes. 
The predicted gate fidelities are above $99.9\%$, over a  wide parameter regime, even in the presence of phonon-induced dephasing, which we treat phenomenologically here.

Moving forward, we note that our analytical formalism, based on a cumulant expansion, can be readily generalized to systems with multiple qubits.
For example, quantum dot spin qubits~\cite{Zajaceaao5965,Watson2018,Yoneda2018}, single-triplet qubits~\cite{Nichol2017}, charge qubits~\cite{PhysRevLett.91.226804, PhysRevLett.95.090502, Kim2015, Yang:2019}, and hybrid qubits~\cite{KimWardSimmonsEtAl2015}, as well as one and two-qubit gate operations, may all be investigated using the methods described here.
In each case, the gate performance can be improved by identifying optimal working points or working strategies, such that the device is less susceptible to the dominant decoherence channel, or other control errors.

\textit{Acknowledgments: } We thank M.\ A.\ Eriksson for enlightening conversations. 
This work was supported in part by ARO (W911NF-17-1-0274) and the Vannevar Bush Faculty Fellowship program sponsored by the Basic Research Office of the Assistant Secretary of Defense for Research and Engineering and funded by the Office of Naval Research through grant N00014-15-1-0029.
The views and conclusions contained in this document are those of the authors and should not be interpreted as representing the official policies, either expressed or implied, of the U.S. Government. The U.S. Government is authorized to reproduce and distribute reprints for Government purposes notwithstanding any copyright notation herein.

\appendix


\section{Details of the Numerical Simulations \label{Sec:Simulation}}
For all the numerical simulations shown in the main text, we follow the method described in Ref.~\cite{Yang:2019}, which we briefly summarize as the follows:
We first generate noise realizations $\{ \delta \varepsilon^{\alpha}(t)\}_{\alpha = 1,\ldots,\alpha_\text{max}}$ whose power spectral density is $1/f$, as described in Eq.~(\ref{Eq:Spectrum One-Over-f}).
We then solve the Schr$\ddot{\text{o}}$dinger equation $i \hbar \frac{d}{dt} |\psi(t)\rangle = [\Hamiltonian_q + \Hamiltonian_\text{ac} +  \Hamiltonian_n] |\psi(t)\rangle $ numerically for each realization using the method we developed in Ref.~\cite{Yang:2019}. 
The solutions are labeled as $\rho^{\alpha} = |\psi^{\alpha}\rangle\langle\psi^{\alpha}|$.
Finally, we evaluate $\langle \rho \rangle$ by averaging over these solutions, $\langle \rho \rangle = \frac{1}{\alpha_\text{max}}\sum\limits_{\alpha = 1}^{\alpha_\text{max}}\rho^{\alpha}$, where for this work, we take $\alpha_\text{max}=10,000$.

\begin{widetext}
\section{Evolution of the Hybrid Qubit in the Absence of Charge Noise}
\label{Sec:AnalyticsNoNoise}

In  this Appendix, we sketch out the perturbative derivation of the time evolution of a strongly driven hybrid qubit in the absence of charge noise, which is in principle exact, when keeping all orders of the expansion.
While portions of these results were first derived in Ref.~\cite{PhysRevA.95.062321}, we present them again here for completeness, because they will be used, below, in our cumulant expansion analysis.

In the absence of noise, the dynamics is governed by the Schr\"{o}dinger equation 
\begin{equation}
i\hbar \frac{d}{dt} U_0 = [\bar \Hamiltonian_q + \bar \Hamiltonian_\text{ac}] U_0.
\end{equation}
Expressing Eq.~(\ref{Eq:HamiltonianQ}) in its eigenbasis $\{|0\rangle,|1\rangle,|L\rangle\}$, we have $\bar \Hamiltonian_q = \text{diag}[E_0,E_1,E_L]$, where $\{E_i\}$ are the energy eigenvalues, and the AC driving Hamiltonian is given by $\bar \Hamiltonian_\text{ac} = \bar V \cos(\omega_d t)$.
In the large-detuning regime, $\varepsilon \gg E_\text{ST},\Delta_1,\Delta_2$, $\bar V$ is given by
\begin{eqnarray}
\label{eq:asymp_driving}
&& \bar V \approx \\
&& A_{\Delta} \begin{pmatrix}
-\frac{\Delta_{1} }{\varepsilon} \!+ \!\frac{\Delta_{2} r}{\varepsilon -E_\text{ST}} & \frac{\Delta_{1} r}{\varepsilon} \!+\!\frac{\Delta_{2}}{\varepsilon-E_\text{ST}} & 1  \\
\frac{\Delta_{1} r}{\varepsilon} \!+\!\frac{\Delta_{2}}{\varepsilon-E_\text{ST}} & \frac{\Delta_{1}}{\varepsilon} \!- \!\frac{\Delta_{2} r}{\varepsilon -E_\text{ST}} &  -r \\
1  & -r & \frac{3 \Delta_{1} }{\varepsilon}\!+\!\frac{3 \Delta_{2} r}{\varepsilon -E_\text{ST}}
\end{pmatrix} 
+
A_{\varepsilon}\begin{pmatrix}
\frac{\Delta_1^2}{2\varepsilon^2}\!-\!\frac{\Delta_2^2}{2(\varepsilon-E_\text{ST})^2} & -\frac{\Delta_{1} \Delta_{2}}{\varepsilon (\varepsilon-E_\text{ST})} & -\frac{\Delta_{1}}{\varepsilon}  \\
-\frac{\Delta_{1} \Delta_{2}}{\varepsilon (\varepsilon-E_\text{ST})} &  -\frac{\Delta_1^2}{2\varepsilon^2}\!+\!\frac{\Delta_2^2}{2(\varepsilon-E_\text{ST})^2} &  \frac{\Delta_{2}}{\varepsilon-E_\text{ST}} \\
-\frac{\Delta_{1}}{\varepsilon}  & \frac{\Delta_{2}}{\varepsilon-E_\text{ST}} & 1 -\frac{3\Delta_1^2}{2\varepsilon^2}-\frac{3\Delta_2^2}{2(\varepsilon-E_\text{ST})^2}
\end{pmatrix} . \nonumber
\end{eqnarray} 
\end{widetext}

We derive $U_0$ using the dressed-state approach of Ref.~\cite{PhysRevA.95.062321}.
We first extend the semiclassical Hamiltonian, $\Hamiltonian_\text{semi} = \bar \Hamiltonian_q + \bar \Hamiltonian_\text{ac} \cos(\omega_d t)$, into a quantum Hamiltonian $\Hamiltonian_\text{QM}$ by imposing the condition
\begin{equation}
\bra{ \alpha } \Hamiltonian_\text{QM}\ket{ \alpha} = \Hamiltonian_\text{semi} + N \hbar\omega_d,
\label{Eq:Correspondence}
\end{equation}
where the coherent state $|\alpha\rangle$ is defined as $a|\alpha\rangle = e^{-i \omega_d\!t} \alpha_0 |\alpha\rangle$, with $a$ the photon annihilation operator and $N = \bra{ \alpha }a^{\dagger} a\ket{\alpha}$ the number of photons.
The quantum Hamiltonian can be written as $\Hamiltonian_\text{QM}=\Hamiltonian_\text{dot}+\Hamiltonian_\text{ph}+V_\text{int}$.
Here, the uncoupled dot Hamiltonian is given by $\Hamiltonian_\text{dot} = \sum_{i=0,1, L} E_i |i\rangle\langle i| \otimes I_\text{ph}$, the uncoupled photon Hamiltonian is given by $\Hamiltonian_\text{ph} = I_\text{dot} \otimes \hbar\omega_d a^{\dagger} a$, and the interaction term is defined as $V_\text{int} = \bar V/2 \sqrt{N}\otimes (a^{\dagger} + a)$.
By evaluating the dynamics of the quantum Hamiltonian, $U_\text{QM}(t) = e^{-\frac{i}{\hbar}\Hamiltonian_\text{QM} t}$, the time evolution of the semiclassical Hamiltonian can be obtained as $U_0(t) = \langle \alpha (t)|U_\text{QM}(t)|\alpha(0)\rangle$.

To simplify the calculation, we note that predominant modes of coherent states occur in the range $n\in [N - \Delta N,N + \Delta N]$, where $\Delta N/N \ll 1$. 
We therefore express the quantum Hamiltonian in the basis of $|i,n\rangle$ for qubit state $|i\rangle$ and photon state $|n\rangle$ for $n\in [N - \Delta N,N + \Delta N]$.
In this way, we obtain
\begin{gather}
\langle i, n |\Hamiltonian_\text{dot}|j, m \rangle = E_i\,\delta_{i,j}\delta_{n,m} , \\
\langle i, n |\Hamiltonian_\text{ph}|j, m \rangle = n \hbar\omega_d \, \delta_{i,j}\delta_{n,m} , \\
\langle i, n |V_\text{int}|j, m\rangle  = \frac{\bar V_{ij}}{2} (\delta_{n,m+1}+ \delta_{n,m-1}).
\end{gather}

To increase the accuracy of $U_0$ when $A_{\varepsilon}$ is large [e.g. the rightmost part of Fig.~3(d) in the main text], we make the following modification in the perturbation:
We first diagonalize the leakage subspace, i.e. the $\langle L, n |\Hamiltonian_\text{QM}|L, m \rangle$ part of the Hamiltonian. 
The new basis now is $\{|0, n\rangle, |1,n\rangle, |\tilde L, n\rangle \}$ where $|\tilde L, n\rangle = \sum_{m=-\infty}^{\infty} J_m(-[\bar V]_{LL}/\hbar \omega_d)|L,n+m\rangle$, and $J_m(x)$ is a Bessel function of the first kind. 
Rewriting the Hamiltonian in this basis, the magnitude of the off-diagonal elements $\tilde V_{ij}$ is now at most ${\cal O}[A_{\varepsilon}\,J_m(-A_{\varepsilon}/\hbar \omega_d)]$.
We then proceed to evalulate $U_0$ perturbatively over $\bar V/\hbar \omega_d$ using the method of Ref.~\cite{PhysRevA.95.062321}.
A similar diagonalization scheme has been applied to floquent treatment of strongly driven two-level system, for example in Ref.~\cite{PhysRevLett.115.133601}.
This is equivalent to performing the basis transformation $U_L = \text{diag}[1,1,e^{-\frac{i \bar V_{LL}}{\hbar \omega}\sin(\omega t)}]$ before going into the dressed space, similar to the basis transformation for the two-level system in Ref.~\cite{PhysRevA.75.063414}.
We use this modification when we perform the analytical calculation of fidelity, plotted as Fig.~3(d) and (e) in the main text.

\section{Evolution in the Presence of Charge Noise}
\label{Sec:AnalyticalCalculation}
To incorporate the effects of detuning noise, we first move to the interaction frame.
Here, the equation of motion for the density matrix $\rho^I$ is given by
\begin{equation}
i \hbar \frac{d}{dt} \rho^I = \delta \varepsilon(t) \mathcal{L} \rho^I ,
\end{equation}
where $\rho^I = U_0^{\dagger} \rho U_0$, $\mathcal{L}\rho^I \equiv [h_n^I,\rho^I]$, and $h_n^I = U_0^{\dagger} \bar h_n U_0$. 
The evolution can be evaluated in terms of the cumulant expansion~\cite{doi:10.1143/JPSJ.17.1100}
\begin{widetext}
\begin{equation}
\langle \rho^I(t) \rangle = 
\exp \left\{ \sum\limits_{n=1}^{\infty} \frac{(-i)^n}{\hbar^n} \int\limits_0^t dt_1 \cdots \int\limits_0^{t_{n-1}} dt_n  \langle \mathcal{L}(t_1) \delta\varepsilon(t_1) \cdots \mathcal{L}(t_n) \delta\varepsilon(t_n) \rangle_c \right\} \rho^I(0) ,
\end{equation}
where $\langle \cdots \rangle$ is the ensemble average over $\delta\varepsilon(t)$ and $\langle \cdots \rangle_c$ is the cumulant average.
To second order in $\delta \varepsilon$, this gives Eq.~(\ref{Eq:AnalyticFormula2nd}) in the main text.

To further simplify the calculation, we express the density and noise matrices in terms of the 3$\times$3 Gellman matrices,  $\{\lambda_i\}_{i=1,\cdots,8}$, which are presented in Appendix.~\ref{Sec:GellmanMatrix}.
The Gellman matrices can be categorized into two subsets: the first set  $\Lambda_q = \{\lambda_1,\lambda_2,\lambda_3,\lambda_4\}$ does not mix qubit and leakage states, while the second set $\Lambda_\text{sup} = \{\lambda_5,\lambda_6,\lambda_7,\lambda_8\}$ mixes the two subspaces. 
Using these matrices, we can write $\rho^I = 1/3 I_3 +1/2 \sum\limits_{i=1}^8 \vec{r}^I_i \lambda_i$ where $I_3$ is the 3$\times$3 identity matrix, and $\vec{r}^I$ is the generalized Bloch vector.
Similarly, we can express the noise matrix in the interaction frame as $h_n^I = \sum\limits_i h_{n,i}^I \lambda_i$.
As noted in the main text, Eq.~(\ref{Eq:AnalyticFormula2nd}) may then be replaced by Eq.~(\ref{Eq:rI(t)}), where
\begin{eqnarray}
\label{Eq:App_K(t)}
[K(t)]_{ij} &=& -\frac{1}{\hbar^2} \int\limits_0^t dt_1 \int\limits_0^{t_1} dt_2 [\sum\limits_{k,l,m = 1}^{8}h_{n,k}^I(t_1)h_{n,l}^I(t_2) T_{k m}^{(i)}T_{lj}^{(m)}] S(t_1-t_2), \nonumber \\
&=& -\frac{1}{\hbar^2} \sum\limits_{\omega_1, \omega_2}
[\alpha_{k,\omega_1}\alpha_{l,\omega_2}  T_{k m}^{(i)}T_{lj}^{(m)} ] I(t,\omega_1,\omega_2). \nonumber
\end{eqnarray} 
Determining the cumulant $K(t)$ therefore reduces to evaluating $I(t,\omega_1,\omega_2)$ in Eq.~(\ref{Eq:I(t)}). 
This integral can be solved analytically for a generic noise spectrum, leading to a complete analytical result for the dynamics that provides a rather accurate description.
It can also be approximated in the asymptotic limit, as in Appendix~\ref{Sec:AnalyticsNoise}, yielding results that the capture key features of the decay profile and the process fidelity.

\section{Gellman Matrices\label{Sec:GellmanMatrix}}
For completeness, we reproduce here the Gellman matrices, $\{\lambda\}_{i=1,\cdots,8}$, which are linearly independent, traceless, and Hermitian, as originally defined in Ref.~\cite{PhysRev.125.1067}:
\begin{equation*}
\begin{array}{llll}
\lambda_1 = 
\begin{pmatrix}
0 & 1 & 0 \\
1 & 0 & 0 \\
0 & 0 & 0
\end{pmatrix},
& 
\lambda_2 = 
\begin{pmatrix}
0 & -i & 0 \\
i & 0 & 0 \\
0 & 0 & 0
\end{pmatrix},
&
\lambda_3 = 
\begin{pmatrix}
1 & 0 & 0 \\
0 & -1 & 0 \\
0 & 0 & 0
\end{pmatrix}, 
&
\lambda_4 = 
\frac{1}{\sqrt{3}}
\begin{pmatrix}
1 & 0 & 0 \\
0 & 1 & 0 \\
0 & 0 & -2
\end{pmatrix},
\\ \\
\lambda_5 = 
\begin{pmatrix}
0 & 0 & 1 \\
0 & 0 & 0 \\
1 & 0 & 0
\end{pmatrix},
&
\lambda_6 = 
\begin{pmatrix}
0 & 0 & -i \\
0 & 0 & 0 \\
i & 0 & 0
\end{pmatrix},
&
\lambda_7 = 
\begin{pmatrix}
0 & 0 & 0 \\
0 & 0 & 1 \\
0 & 1 & 0
\end{pmatrix},
&
\lambda_8 = 
\begin{pmatrix}
0 & 0 & 0 \\
0 & 0 & -i \\
0 & i & 0
\end{pmatrix}.
\end{array}
\end{equation*}

\section{Asymptotic Results in the Long-Time Limit}
\label{Sec:AnalyticsNoise}
In the presence of noise, the dynamics of the density matrix in the interaction frame is governed by the Schr$\ddot{\text{o}}$dinger equation 
\begin{equation}
i\hbar \frac{d}{dt} \rho^I = \delta \varepsilon \mathcal{L}\rho^I \equiv \delta \varepsilon [h_n^I,  \rho^I].
\end{equation}
In the large detuning regime, the noise matrix can be approximated (by expanding in powers of the tunnel couplings) as 
\begin{equation}
\bar{h}_n \simeq 
\begin{pmatrix}
\frac{\Delta_1^2}{2\varepsilon^2}\!-\!\frac{\Delta_2^2}{2(\varepsilon-E_\text{ST})^2} & -\frac{\Delta_{1} \Delta_{2}}{\varepsilon (\varepsilon-E_\text{ST})} & -\frac{\Delta_{1}}{\varepsilon}  \\
-\frac{\Delta_{1} \Delta_{2}}{\varepsilon (\varepsilon-E_\text{ST})} &  -\frac{\Delta_1^2}{2\varepsilon^2}\!+\!\frac{\Delta_2^2}{2(\varepsilon-E_\text{ST})^2} &  \frac{\Delta_{2}}{\varepsilon-E_\text{ST}} \\
-\frac{\Delta_{1}}{\varepsilon}  & \frac{\Delta_{2}}{\varepsilon-E_\text{ST}} & 1 -\frac{3\Delta_1^2}{2\varepsilon^2}-\frac{3\Delta_2^2}{2(\varepsilon-E_\text{ST})^2}
\end{pmatrix} ,
\end{equation}
similar to the second term of Eq.~(\ref{eq:asymp_driving}).
We note that, by choosing $\Delta_1 = \Delta_2 \sim 0.7 \,E_\text{ST}$, as noted in the main text, the main term controlling the dephasing is $\bar{h}_{n,3} \approx 0$, consistent with a higher-order DC sweet spot~\cite{Wong2016, PhysRevA.95.062321}.

In this work, we focus on detuning charge noise with a $1/f$ power spectral density, as defined in Eq.~(\ref{Eq:Spectrum One-Over-f}).
The corresponding time correlation function is given by
\begin{equation}
S(t) = 2 c_{\varepsilon}^2 \left[ \Ci (\omega_h |t|) - \Ci (\omega_l |t|) \right],
\end{equation}
where $\Ci(x)$ is the cosine integral.
For this case, we derive the the asymptotic results, Eqs.~(10) and (11), discussed in the main text, as follows.
\begin{enumerate}[label = {(\arabic*)}]
	\item First, calculate the Fourier components $\alpha_{i,\omega}$, as defined in $h_{n}^I(t) = U_0^{\dagger} \bar h_n U_0=\sum\limits_{i,\omega}\alpha_{i,\omega} e^{i\omega t}\lambda_i$, and insert these results into Eq.~(\ref{Eq:K(t)}).  
	\item Next, approximate $I(t,\omega_1,\omega_2)$ [Eq.~(\ref{Eq:I(t)})] in the regime of interest \yc{and only retain non-oscillatory terms.} 
	(a) For the decay profile, Eq.~(\ref{Eq:AsymptoticFormula}), we consider the long-time limit, $1/\omega_l \gg t/(2\pi) \gg 1/\omega \gg 1/\omega_h$, where $\omega$ is any angular frequency relevant to the hybrid qubit. 
	(b) For the gate infidelity in the rotating frame, Eq.~(\ref{Eq:FidelityIAsymptotic}), we consider the limit  $1/\omega_l \gg t/(2\pi)\sim 1/\Omega \gg 1/\omega \gg 1/\omega_h$, where $\Omega$ is the Rabi angular frequency.
	\item Finally, evaluate $K(t)$ in Eq.~(\ref{Eq:K(t)}).
	In the long-time limit, the dominant contribution to $K$ arises from pure dephasing in the rotating frame, $K_{\varphi} = D_{\varphi} I(t,\omega_1 = 0,\omega_2 =0)$, where $D_{\varphi}$ is an 8$\times$8 matrix of coefficients.
	To evaluate \yc{$\exp[K(t)]$}, we first perform a Schrieffer-Wolff decomposition \yc{$D_{\varphi}$}, up to ${\cal O}[\gamma^2]$, to decouple the qubit and leakage subspaces.
	The decoupled subspaces exhibit distinct behaviors: the qubit subspace exhibits slow dephasing, with $[D_{\varphi}]_{ij}\sim {\cal O}[\gamma^2]$, while the leakage subspace exhibits fast dephasing with $[D_{\varphi}]_{ij}\sim {\cal O}[\gamma^0]$.
	Consequently, the elements in $e^{K_{\varphi}}$ all vanish at this order, except in the space spanned by $\Lambda_q$.
	Similarly, we evaluate the leading order terms in $K_\text{M}$ and $K_{\text{nMn}\varphi}$, which occur at ${\cal O}[\gamma^0]$.
	For pedagogical purposes, we retain the leading order terms in each of these categories, and evaluate the exponential $e^{K(t)} = e^{K_{\varphi}(t) + K_\text{M}(t)+K_{\text{nMn}\varphi}(t)}$.
\end{enumerate}

The final asymptotic results are summarized as the follows.
\begin{enumerate}[label = {(\alph*)}]
	\item For the time limit appropriate for the decay profile, we obtain
	\begin{equation}
\label{Eq:IntergalOneOverF_LongTime}
I(t,\omega_1,\omega_2) 
\approx \left\{
\begin{array}{lc}
c_{\varepsilon}^2\; t^2 \left[\log(1/\omega_l t) -\gamma_E+ 3/2 \right], & (\omega_1 = \omega_2 = 0), \\
c_{\varepsilon}^2 \; t\, (\pi /| \omega_1|+ 2 i \log|\omega_1/\omega_l|/ \omega_1),  & (\omega_1 = - \omega_2 \neq 0), \\
2 i \; t\,(\log(1/\omega_l t) -\gamma_E+ 1)/\omega_2, & (\omega_1 = 0 \text{ and } \omega_2 \neq 0), \\
0, & (\text{Otherwise}),
\end{array}
\right. 
\end{equation}
where $\gamma_E \approx 0.5772$ is Euler's constant, and
	\begin{eqnarray}
	\label{Eq:AsymptoticFormulaMatrix}
&&[e^{K(t)}]_{1-4,1-4}  \\
&&\approx
\begin{pmatrix}
(1-\frac{a}{3})e^{-\Gamma_x \, t}  & 0 & 0 & b(e^{-\Gamma_x\, t}+e^{- \Gamma_L\, t}) \\
0 & (1-\frac{a}{3})e^{-\Gamma_y\, t-\Gamma_{\varphi}\,\varphi(t)}  & 0 & 0 \\
0 & 0 & (1-\frac{a}{3})e^{-\Gamma_y\, t-\Gamma_{\varphi}\,\varphi(t)} & g(e^{-\Gamma_y\, t-\Gamma_{\varphi}\,\varphi(t)}+e^{-\Gamma_L\, t}) \\
b(e^{-\Gamma_x\, t}+e^{-\Gamma_L\, t}) & 0 & g(e^{-\Gamma_y\, t-\Gamma_{\varphi}\,\varphi(t)}+e^{-\Gamma_L\, t}) & (1-a)e^{-\Gamma_L\, t} \nonumber \\
\end{pmatrix},
\end{eqnarray}
where $[e^{K(t)}]_{ij}=0$ otherwise.
Here, $\varphi(t)$, $\Gamma_x$, $\Gamma_y$, $\Gamma_L$, $\Gamma_\varphi$, and $a$ are defined in the main text, and we further define
\begin{gather}
b = - \frac{\sqrt{3} (E_L-E_0)(E_L-E_1)\bar V_{0L} \bar V_{1L}}{(E_L-E_0-h\,f)(E_L-E_0 + h\,f)(E_L-E_1 -h\,f)(E_L+E_1 + h\,f)}, 
\label{Eq:b} \\
g = -\frac{3 \sqrt{3}(E_L-E_0)^2 \bar V_{0L}^2}{6 (E_L-E_0-h\,f)^2(E_L-E_0 + h\,f)^2} + \frac{3\sqrt{3} (E_L-E_1)^2 \bar V_{1L}^2}{6 (E_L-E_1 -h\,f)^2(E_L+E_1 + h\,f)^2}.
\end{gather}
In Fig.~\ref{Fig:FigS1}, we plot the resulting figure of merit, $f_\text{Rabi}/\Gamma_\text{Rabi}$, where the Rabi decay rate $\Gamma_\text{Rabi}$ is identified from Eq.~(\ref{Eq:AsymptoticFormulaMatrix}), where we define $\langle \rho^I_{00}\left(t =1/\Gamma_\text{Rabi}\right)\rangle \equiv (1 + 2e^{-1})/3$, for the initial state $\rho^I(t=0)=|0\rangle\langle 0|$.
The white dashed line in the figure is obtained by maximizing the figure of merit, or equivalently, by setting $\Gamma_{\varphi} \sim (d f_\text{Rabi}/d\varepsilon)^2 = 0$, as consistent with an AC sweet spot \cite{2018arXiv180701310D}.  
This condition determines the optimal combination of driving amplitudes, as described in the main text.
	\item For the calculation of the gate infidelity in the rotating frame, we note that the condition $t/(2\pi) \sim 1/\Omega$ makes the calculation somewhat more complicated. 
	Fortunately, the DC sweet-spot condition, $\bar h_{n,3}=0$, causes $\alpha_{i,\pm\Omega} = 0 $, at $O[\gamma^2]$, which simplifies the calculation. 
	After lengthy manipulations, we finally arrive at Eq.~(\ref{Eq:FidelityIAsymptotic}) in the main text.
\end{enumerate}
\end{widetext}

\begin{figure}[t]
\includegraphics[width=3in]{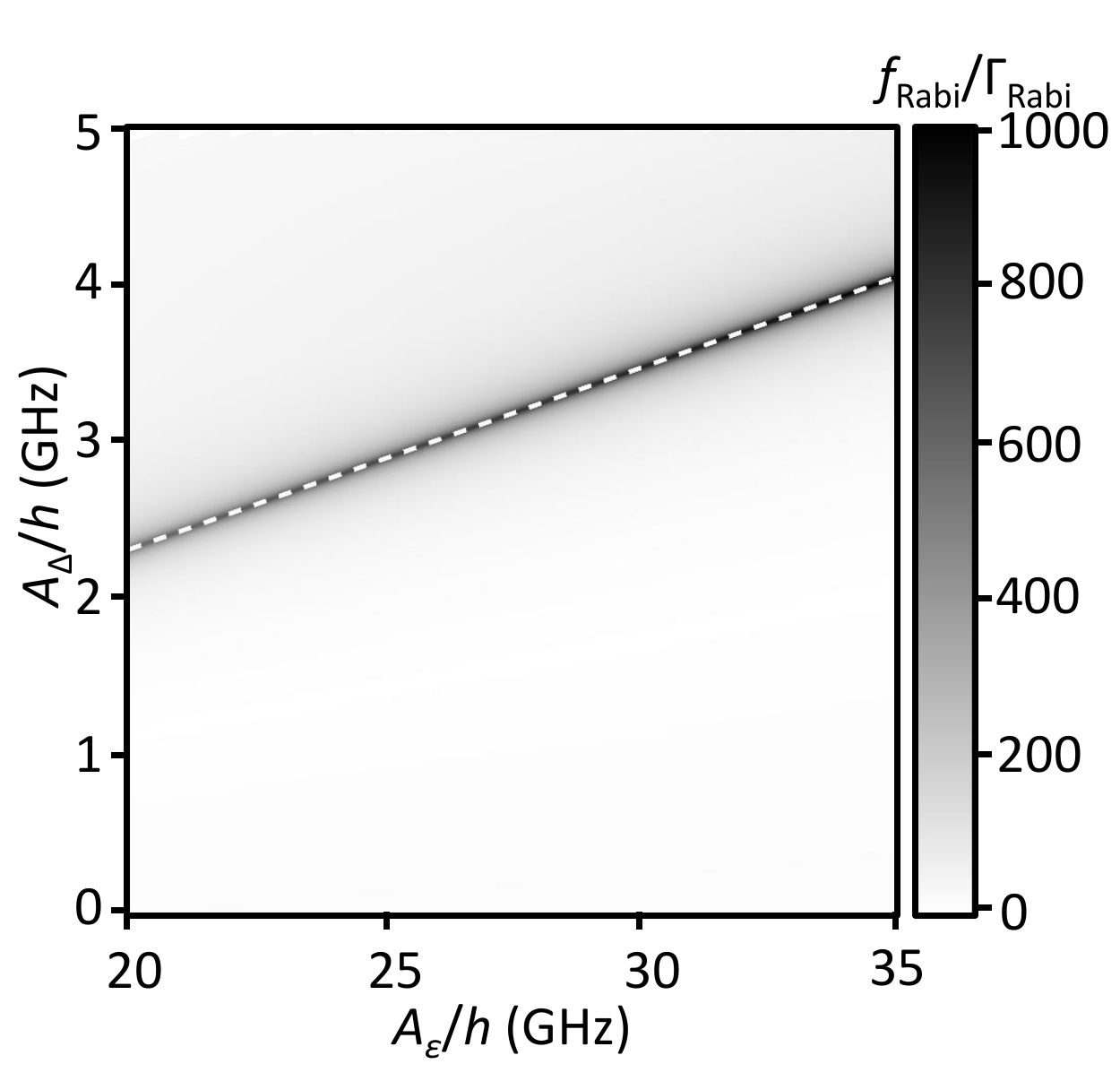}
\caption{
\label{Fig:FigS1}
Simulation results for the figure of merit, $f_\text{Rabi}/\Gamma_\text{Rabi}$, of a strongly driven quantum dot hybrid qubit, as a function of the detuning driving amplitude $A_{\epsilon}$ and the tunnel coupling driving amplitude $A_{\Delta}$, for the same DC control parameters and charge noise parameters used in Figs.~\ref{Fig:Fig2} and \ref{Fig:Fig3} of the main text.
Here, $\Gamma_\text{Rabi} = 1/ T_\text{Rabi}$ is the Rabi decay rate, estimated from Eq.~(\ref{Eq:AsymptoticFormulaMatrix}), and  $f_\text{Rabi}$ is the Rabi frequency. 
The dashed white line corresponds to an AC sweet spot, $d f_\text{Rabi}/d\varepsilon=0$, for which the $\Gamma_{\varphi}$ dephasing term vanishes, up to ${\cal O}[(V/\hbar \omega_d)^2]$, yielding an optimal figure of merit.
This line accurately matches the simulations, and is the same white dashed line plotted in Fig.~\ref{Fig:Fig3}.  
At the AC sweet spot, the figure of merit can be $>$1000, consistent with $X_{\pi}$ fidelities approaching $99.9\%$.
}
\end{figure}

\begin{figure*}[t]
\includegraphics[width=7in]{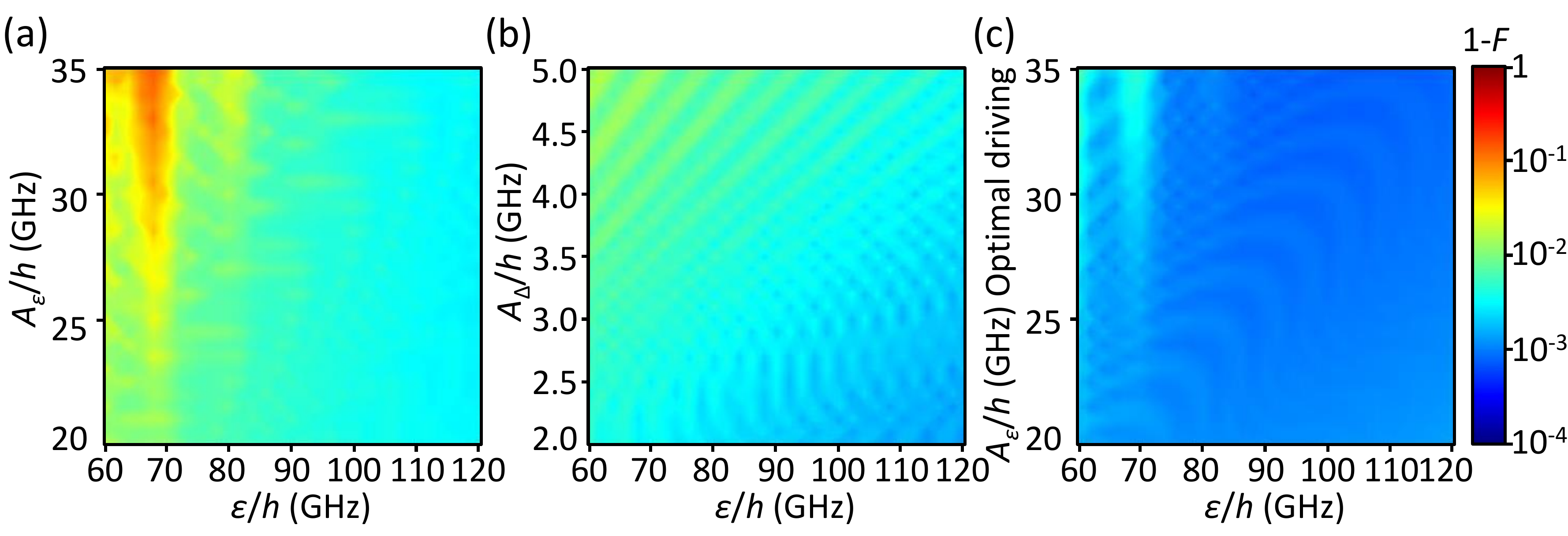}
\caption{
\label{Fig:FigS2}
Comparison of the infidelities of $X_{\pi}$ gates, for three different AC driving methods: detuning driving, tunnel coupling driving, and optimal driving at an AC sweet spot, in the presence of detuning noise and phenomenological phonon dephasing.
The results are presented in the laboratory frame, and assume rectangular pulse envelopes.
We use the same DC control parameters and noise parameters as Figs.~\ref{Fig:Fig2} and \ref{Fig:Fig3}, with a variable detuning parameter $\varepsilon$.
(a) Detuning driving. Here, the vertically oriented, low-fidelity feature is caused by leakage. 
The fidelity attains a maximum value of $F = 0.9973 $ at the point $\{\varepsilon, A_{\epsilon}\}/h = \{120, 27\} \; \units{GHz}$.
(b) Tunnel-coupling driving. 
The fidelity attains a maximum value of  $F = 0.9989 $ at the point $\{\varepsilon, A_{\Delta}\}/h = \{120, 2\} \; \units{GHz}$.
(c) Optimal, simulatanteous driving of the detuning and tunnel coupling parameters at an AC sweet spot.
The fidelity is higher than in (a) and (b) for almost every parameter value, and attains a maximum value $F = 0.9993 $ at the point $\{\varepsilon, A_{\epsilon},A_{\Delta}\}/h = \{88, 35, 3.64\} \; \units{GHz}$. 
Importantly, we observe a broad range of parameters where $F>0.999$, particularly when $\varepsilon\gtrsim 90$~GHz.
}
\end{figure*}

\section{Process Fidelity}
\label{Sec:ProcessFidelity}
Following Ref.~\cite{ChuangBook}, a generic quantum process $\mathcal{E}$ on a three-dimensional Hilbert space may be expressed as
\begin{equation}
\mathcal{E}(\rho_0) = \sum\limits_{m,n} E_m \rho_0 E_n ^{\dagger} \chi_{mn},
\end{equation}
where $\{E_m\}$ is a basis for the vector space of $3\times 3$ matrices, $\chi$ is the process matrix, and $\rho_0$ represents any initial density matrix.
The process fidelity is then defined as $F=\text{Tr} [\chi_\text{sys} \chi_\text{ideal}]$, where $\chi_\text{sys}$ is the actual process matrix, describing the system evolution, including non-ideal contributions from strong driving and decoherence, and $\chi_\text{ideal}$ describes the ideal evolution.
Since $\chi_\text{ideal}$ does not involve the leakage channel, we simplify the calculation of $F$ by projecting $\rho_0$ and $\mathcal{E}$ onto the 2D logical subspace and solving for the corresponding 4$\times$4 $\chi$ matrices~\cite{PhysRevA.95.062321}. 
In this case, we choose $E_m$ from the Pauli basis $\{I, \sigma_x, -i \sigma_y, \sigma_z\}$ and follow the standard procedure for computing $F$~\cite{ChuangBook}.
For an $X_{\pi}$ gate, the ideal operation is defined as 
\begin{equation}
U_\text{ideal} = 
\begin{pmatrix}
0 & - i e^{ - i \tilde{E}_0 t_g/\hbar}\\
-i e^{ - i\tilde{E}_1 t_g/\hbar} & 0
\end{pmatrix} 
\end{equation}
where $\tilde{E}_0$ and $\tilde{E}_1$ are the qubit energies renormalized by the Bloch-Siegert shifts. 

By moving to the interaction frame, defined by $U_0$, we can focus specifically on the effects of decoherence on the fidelity, $F^I$.
Here, we compare the $\chi$ matrix of the actual process, $\chi^I_\text{sys}$, with the ideal process, defined as $\chi^I_\text{ideal} = \text{diag}[1,0,0,0]$:
\begin{equation}
F^I = \text{Tr} [\chi^I_\text{sys} \chi^I_\text{ideal}].
\end{equation}
As noted in the main text, $\chi^I_\text{sys}$ is not affected by strong driving effects, which are already included in $U_0$.
$F^I$ therefore describes the fidelity due to decoherence (only), and serves as an upper-bound on the  fidelity in the laboratory frame, where $F$ is suppressed by strong driving effects as well as decoherence.

\section{Comparison Of Different Drives \label{Sec:ComparisonDrive}}
In this section, we compare the performance of three different methods of AC driving:  detuning driving, tunnel coupling driving, and simultaneous driving of the detuning and the tunnel coupling, along the AC sweet spot (dashed line) in Fig.~\ref{Fig:FigS1}.
In each case, we numerically simulate the $X_{\pi}$ gate fidelity in the presence of $1/f$ detuning charge noise and phonon effects, introduced as in Eq.~(\ref{Eq:FidelityPhonon}) of the main text.
Here, we take $T_\text{ph} = 5\; \units{\mu s}$ as the phonon decay rate.
In Fig.~\ref{Fig:FigS2}, we plot the resulting infidelities as a function of detuning parameter, $\varepsilon$, and the appropriate amplitude for each type of drive. 
We find that the optimal, combined drive is  superior to the other options over the entire parameter space.
Moreover, the optimal drive exhibits a broad regime of parameters for which the fidelities are $>$99.9\%.

\begin{figure*}[t]
\includegraphics[width=3in]{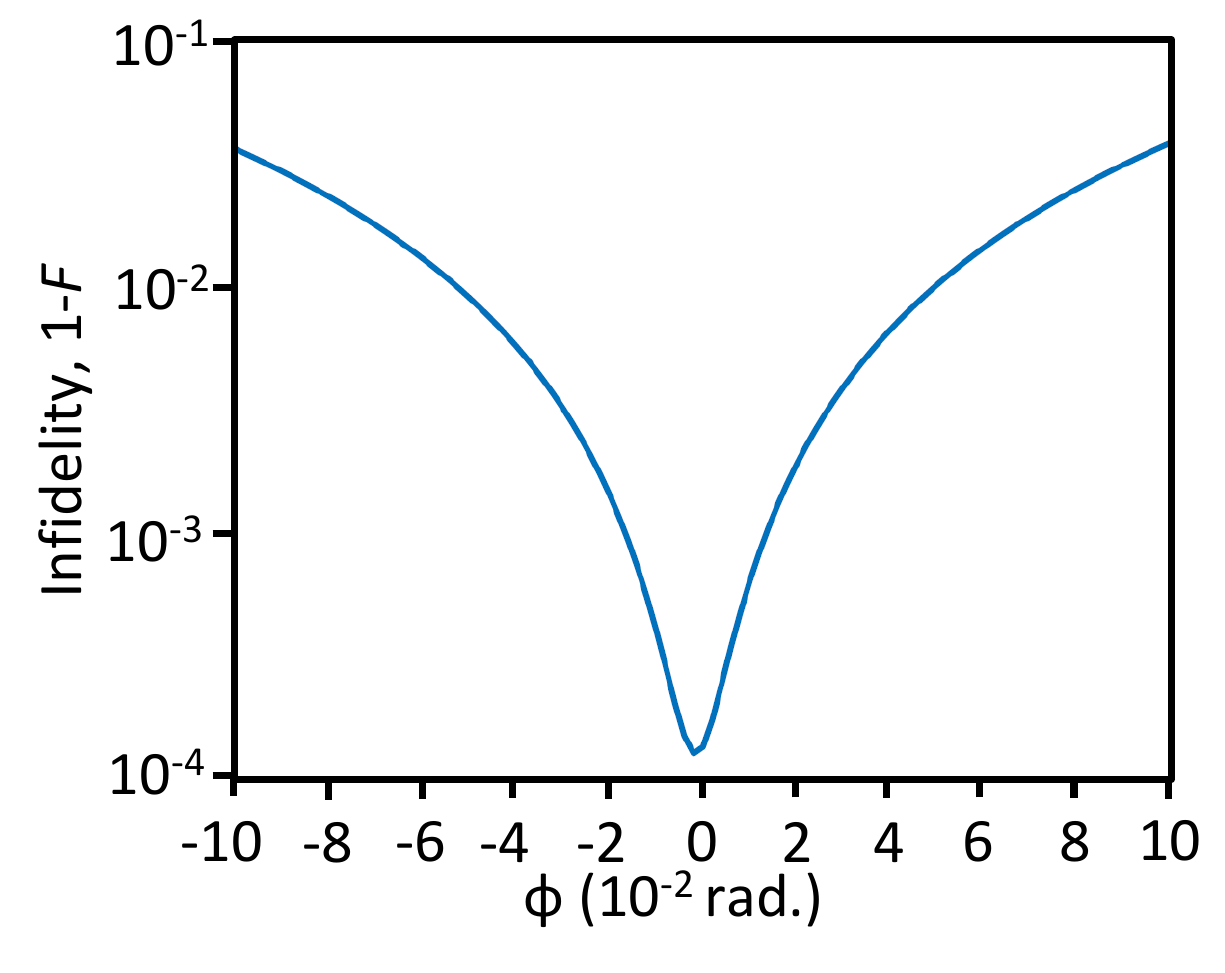}
\caption{
\label{Fig:FigS3}
Dependence of the infidelity on the phase shift between the detuning and tunnel coupling drives.
Here the hybrid qubit is driven at an AC sweet spot, with rectangular pulse envelopes.
The DC control parameters are the same as in Figs.~\ref{Fig:Fig2} and \ref{Fig:Fig3}, while the AC parameters are $\{A_\varepsilon,A_\Delta\}/h = \{27,  3.1\} \,\units{GHz}$.
For the range of phase shift shown here, the infidelity is found to vary by more than two orders of magnitude.
To suppress this effect below $0.1\%$, the phase shift should be less than $0.02 \,\units{rad.} \approx 1^{\circ}$.
}
\end{figure*}

\section{Infidelity Caused by Phase Errors \label{Sec:PhaseError}}
For the optimal driving method, we now show that the infidelity is sensitive to the relative phase between the two driving terms, even when the driving amplitudes are tuned to an AC sweet spot.
In our simulations, we include a variable phase shift $\phi$ to the tunnel coupling driving term in the modified driving Hamiltonian, given by
\begin{widetext}
\begin{equation}
\Hamiltonian_\text{ac} =
\begin{pmatrix}
-\frac{A_{\varepsilon} \cos(\omega_d t)}{2} & 0 & A_{\Delta}\cos(\omega_d t + \phi) \\
0 & -\frac{A_{\varepsilon} \cos(\omega_d t)}{2}  & -r A_{\Delta} \cos(\omega_d t + \phi) \\
A_{\Delta}\cos(\omega_d t + \phi)  & -r A_{\Delta}\cos(\omega_d t + \phi)  & \frac{A_{\varepsilon}\cos(\omega_d t)}{2}
\end{pmatrix}.
\label{Eq:HamiltonianAC}
\end{equation}
\end{widetext}
[Compare to Eq.~(\ref{Eq:Hamiltonianac}).]
Figure~\ref{Fig:FigS3} shows the simulated dependence of the infidelity on $\phi$.
As a benchmark, for this example, we find that achieving an infidelity below $0.1\%$ requires phases errors below $0.02 \,\units{rad.} \approx 1^{\circ}$.

\bibliography{DQHQNoise.bib}

\end{document}